\documentstyle[11pt,newpasp,epsfig,twoside]{article}
\def\edcomment#1{\iffalse\marginpar{\raggedright\sl#1\/}\else\relax\fi}
\reversemarginpar

\begin{document}
\title{Resonant Gaps in the Scattered Cometary Population\\ of the
 Trans-Neptunian Region}

\author{\bf Tanya Taidakova}

\affil{Crimean Astrophysical Observatory, Ukraine}

\author{\bf Leonid M. Ozernoy}

\affil{5C3, School of Computational Sciences and Department of Physics
\& Astronomy, George Mason U., Fairfax, VA 22030-4444, USA}

\author{\bf Nick N. Gorkavyi}

\affil{NRC/NAS; NASA/GSFC, Greenbelt, MD 20771, USA}

\begin{abstract}
Our numerical simulations of the Edgeworth-Kuiper belt objects
gravitationally scattered by the four giant planets accounting for mean motion 
resonances reveal numerous resonant gaps in the distribution of the scattered
population.
\end{abstract}

Available numerical simulations
indicate that gravitational scattering of the EKBOs by the four giant
planets might explain the transport of comets from the trans-Neptunian
region all the way inward, down to Jupiter, as well as the origin of
the so called scattered disk (for a review, see Malhotra et al. 1999).
Our recent simulations exploring both the spatial and phase structure of
cometary populations beyond Jupiter (Ozernoy, Gorkavyi, \& Taidakova 2000
$\equiv$ OGT)  deal with 36 test bodies, initial positions of which were
taken close to the known EKBOs so that their orbits intersected the 
Neptune's orbit. Positions of every of the scattered  test bodies were 
traced for $\sim 0.5$~Gyrs and recorded each revolution of Neptune about the 
Sun, {\it e.g.} 165 yrs, totalling $\sim 2.8\cdot 10^7$ positions. 
This delineates in great detail the spatial as well as phase structure of 
the cometary population beyond Jupiter.

Our simulations
reveal that (i) each giant planet dynamically controls a cometary population
called the ``cometary belt" (a well-known `cometary family', such as the
Jupiter family comets, is just a visible part of the cometary belt); and
(ii) comets avoid resonant orbits so that the belts contain
numerous gaps in the $(a,e)$- and $(a,i)$-spaces similar to the
Kirkwood gaps in the main asteroidal belt. We have found that a numerous
resonant gap structure is mantained in the scattered cometary population
far beyond the Neptunian belt.

Fig.~1 shows how the distribution of the scattered bodies
in semimajor axis, $a$, evolves on a time scale of 0.5 Gyrs. Every bin of
$\Delta a=0.1$ AU was visited by each of the test bodies, on average, $\sim
500$ times, which confirms a strong level of chaos in the scattered
population. Our following findings are worth mentioning:

(i) Numerous gaps at the resonances with Neptune are well pronounced;
(ii) A robust resonant structure of gaps is rapidly established;
(iii) Unlike to 
\begin{figure} [!ht]
\centerline{\epsfig{file=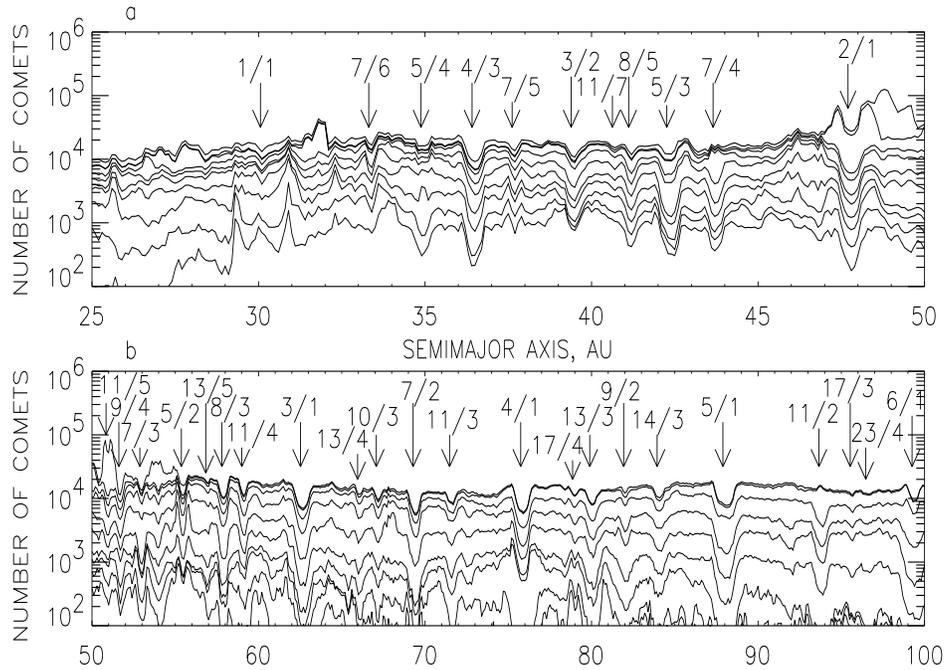,width=5.25in,height=4.44in}} 
\caption{
Evolution of the distribution of simulated scattered comets 
in semimajor axis: 
{\bf a} for comets with $25<a<50$ AU
and {\bf b} for comets with $50<a<100$ AU. Number of comets is given
between $a$ and $a+\Delta a$, where $\Delta a=0.1$ AU.
Arrows indicate various resonances with Neptune. 
Ten distributions, from below to the top, are shown at
1, 2, 4, 8, 16, 32, 64, 128, 256, and 500 Myrs, respectively.
}
\label{fig1}
\end{figure}
\noindent
the distribution of the scattered objects in heliocentric
distance, which has a maximum near the Neptunian orbit (Levison \& Duncan 
1997, OGT), the distribution in semimajor axis appears rather uniform;
(iv) There are appreciable accumulations of the scattered comets near the
resonances well outside the zone of a strong gravitational scattering
({\it e.g.}, near the 2:1 resonance). Those accumulations (called `diffusive' 
ones in OGT) are produced by slowing down of diffusion as the comet's
distance of pericenter turns out to be outside the planetary orbit.

\end{document}